\documentclass[11pt]{article}
\usepackage{lmodern}
\usepackage{fourier}
\usepackage{setspace}

\PassOptionsToPackage{hyphens}{url}

\usepackage[letterpaper,top=0.75in,total={6.5in,9.5in}]{geometry}

\usepackage{microtype}
\UseMicrotypeSet[protrusion]{basicmath} % disable protrusion for tt fonts

\usepackage[pdftex,rgb,dvipsnames,svgnames,hyperref,table]{xcolor}
\definecolor{darkblue}{rgb}{0,0.3,0.68}

\usepackage[pdftex,breaklinks=true,colorlinks=true,
bookmarks=false,pdfhighlight=/O,urlcolor=darkblue,
linkcolor=darkblue,citecolor=mediumgreen]{hyperref}

\usepackage{titlesec}
\titlespacing\section{0pt}{8pt plus 4pt minus 2pt}{0pt plus 2pt minus 2pt}
\titlespacing\subsection{0pt}{8pt plus 4pt minus 2pt}{0pt plus 2pt minus 2pt}
\titlespacing\subsubsection{0pt}{8pt plus 4pt minus 2pt}{0pt plus 2pt minus 2pt}
\titleformat{\section}{\normalfont\Large\bfseries}{}{0pt}{}
\titleformat{\subsection}{\slshape\large\bfseries}{}{0pt}{}

\usepackage{tocloft}
\makeatletter
\renewcommand{\cftsecpresnum}{\begin{lrbox}{\@tempboxa}}
\renewcommand{\cftsecaftersnum}{\end{lrbox}}
\renewcommand{\cftsubsecpresnum}{\begin{lrbox}{\@tempboxa}}
\renewcommand{\cftsubsecaftersnum}{\end{lrbox}}
\makeatother

\title{\textls[-27]{Nine Best Practices for Research Software Registries and Repositories:}\\
A Concise Guide}

\date{23 December 2020}

\author{Task Force on Best Practices for Software Registries$\,$\footnote{Corresponding authors: \href{mailto:aallen@ascl.net}{Alice Allen}, \href{mailto:mhucka@caltech.edu}{Mike Hucka}, and \href{mailto:tmorrell@library.caltech.edu}{Tom Morrell}.\newline
This Task Force was convened by the \href{https://github.com/force11/force11-sciwg}{\emph{FORCE11 Software Citation Implementation Working Group}}.}}

\begin{document}

\maketitle
\thispagestyle{empty} % Needs to come after \maketitle to work on 1st page.

% Do not use \abstract{}; it changes the format *globally*.  I don't have time
% to debug what's going on.  The abstract is simple enough to do this way:
{\small\vspace*{-10pt} Scientific software registries and repositories serve various roles in their respective disciplines. These resources improve software discoverability and research transparency, provide information for software citations, and foster preservation of computational methods that might otherwise be lost over time, thereby supporting research reproducibility and replicability. However, developing these resources takes effort, and few guidelines are available to help prospective creators of registries and repositories. To address this need, we present a set of nine best practices that can help managers define the scope, practices, and rules that govern individual registries and repositories. These best practices were distilled from the experiences of the creators of existing resources, convened by a Task Force of the \href{https://github.com/force11/force11-sciwg}{\emph{FORCE11 Software Citation Implementation Working Group}} during the years 2019--2020. We believe that putting in place specific policies such as those presented here will help scientific software registries and repositories better serve their users and their disciplines.}

\tableofcontents

% After this point, make each section start on a new page.
\newcommand{\sectionbreak}{\clearpage}

\section{Introduction}
\label{introduction}

Scientific software registries and repositories serve various roles in their respective disciplines. \emph{Registries} are typically indexes or catalogs of software stored elsewhere, while \emph{repositories} are both indexes \emph{and} places where software is stored. Both types of resource improve software discoverability and research transparency, provide information for software citations, and foster preservation of computational methods that might otherwise be lost over time, thereby supporting research reproducibility and replicability. Many provide or are integrated with other services, including indexing and archival services, which can be leveraged by librarians, digital archivists, journal editors and publishers, and researchers alike.

Having specific policies in place for software registries and repositories ensures that users and administrators have reference documents to help define a shared understanding of the scope, practices, and rules that govern these collections. These practices can prove useful in a variety of situations, including, but not limited to, presenting the contents in the resource to stakeholders and community members, reassuring potential contributors by clarifying sensitive issues such as attribution, and defining how content in a registry or repository can be (re)used by others.

The best practices presented here were proposed and developed by a Task Force of the \href{https://github.com/force11/force11-sciwg}{\emph{FORCE11 Software Citation Implementation Working Group}}.  The members of the Task Force were managers and editors of scientific software registries and repositories.  Development of the best practices began with a series of monthly conference calls in 2019 and continued at the \href{https://asclnet.github.io/SWRegistryWorkshop/}{\emph{Scientific Software Registry Collaboration Workshop}}, a two-day workshop held at the University of Maryland in November, 2019, with generous funding from the \href{https://sloan.org}{Sloan Foundation}. In 2020, the Task Force made additional refinements to the best practices during virtual meetings and through online collaborative writing.  The \hyperref[authors]{Authors} section lists the people who participated.

Each guideline is presented below with an explanation as to why we recommend the practice, what the practice describes or contains, and specific considerations to take into account.  Our recommendations are partitioned into nine separate policies or statements, though there is inescapable overlap between some of them. In practice, the statements and policies are often combined into a smaller number of documents, as is evident in most of the real-world examples presented at the end of this document. To reduce repetition in the descriptions of the guidelines, we often refer to registries and repositories collectively as ``resources'' and ``collections.''  

These nine best practices, though not an exhaustive list, are applicable to the varied resources represented in the Task Force, so are likely to be broadly applicable to other scientific software repositories and registries. We believe that adopting these practices will help document, guide, and preserve these resources, and put them in a stronger position to serve their disciplines, users, and communities.

\section{Best Practice: Provide a public scope statement}
\label{best-practice-provide-a-public-scope-statement}

\textbf{Why we recommend this}: A scope statement clarifies the type of software contained in the repository or indexed in the registry. This manages the expectations of the potential depositor of metadata and/or software, as well as the resource seeker. It informs both of what the collection does and does not contain.

\textbf{This should describe}:

\begin{itemize}
\item What is accepted, and acceptable

\item What is not accepted

\item Exceptions to either/both of the above if necessary

\end{itemize}

\textbf{What you might consider when writing a scope statement: }

\begin{itemize}
\item Defining the community being served

\item The types of software listed in the registry or stored in the repository, such as source code or compiled executables

\item Criteria that must be satisfied by accepted software, such as whether certain software quality metrics must be fulfilled or whether the software must be used in published research

\item Whether the code has to be in the public domain and/or have a license from a predefined set

\item Whether software registered in another registry or repository will be accepted

\end{itemize}

\section{Best Practice: Provide guidance for users}
\label{best-practice-provide-guidance-for-users}

\textbf{Why we recommend this}: Different users of the registry or repository will benefit from having guidance on how to access the information they are interested in. For example, it is useful to describe how to search the collection, answer frequently asked questions (FAQs), provide tips and tricks, and to let users know who to contact for assistance.

A separate section in these guidelines on the \hyperref[best-practice-stipulate-conditions-of-use]{\emph{Conditions of use policy}} covers terms of use of the collection, including data and API, and how best to cite records in the resource and the resource itself. Guidance for users who wish to contribute software is covered in the next section, \hyperref[best-practice-provide-guidance-to-software-contributors]{\emph{Provide guidance to software contributors}}.

\textbf{This should describe}:

\begin{itemize}
\item How to perform common user tasks

\item Answers to questions that are often asked or can be anticipated

\item Whom to contact for questions or help

\end{itemize}

\textbf{What you might consider when writing guidance for users:}

\begin{itemize}
\item Identifying the types of users your resource has or could potentially have, and corresponding use cases

\item Offering multiple forms of guidance, such as in-field prompts, linked explanations, and completed examples

\item If there is an API, including a description specifying the interface or a pointer to the official documentation for the interface

\item If content negotiation is enabled, stating what formats, such as JSON-LD or XML, are supported

\end{itemize}

\section{Best Practice: Provide guidance to software contributors}
\label{best-practice-provide-guidance-to-software-contributors}

\textbf{Why we recommend this:} People interested in contributing software entries to the registry or repository need to know what the process entails. The scope statement will already have explained \emph{what} is accepted and what is not; the contributor policy addresses \emph{who} can add or change software entries, and the processes involved.

\textbf{This should describe}:

\begin{itemize}
\item Who can or cannot submit entries and/or metadata

\item Required and optional metadata expected from software contributors

\item Review process, if any

\item Curation process, if any

\item Procedures for updates (e.g., who can do it, when it is done, how is it done)

\end{itemize}

\textbf{What you might consider when writing a contributor policy}:

\begin{itemize}
\item Defining who can submit and/or update entries

\item Whether the author(s) of the software will be contacted if the contributor is not also an author, and whether contact is a condition or side-effect of the submission

\item Stating how persistent identifiers are assigned (if they are used)

\item Including a statement that depositors must comply with all applicable laws and not be intentionally malicious

\end{itemize}

\section{Best Practice: Establish an authorship policy}
\label{best-practice-establish-an-authorship-policy}

\textbf{Why we recommend this}: Establishing a policy dedicated to authorship ensures that people are given due credit for their work. It also serves as a document that administrators can turn to in case authorial disputes arise and allows for proactive problem mitigation, rather than having to resort to reactive interpretation. Further, having an authorship policy is in keeping with similar policies by journals and publishers. Having such explicit authorship policies is thus part of a larger trend. Note that the authorship policy will be communicated at least partially to users through guidance provided to software contributors.

\textbf{This should describe}:

\begin{itemize}
\item Who should be listed as an author of the software

\item Policies around making changes to authorship

\item How authorship disputes are handled

\item What the resource will do in case of conflict

\end{itemize}

\textbf{What you might consider when writing an authorship policy:}

\begin{itemize}
\item Taking into consideration whether those who are not coders, such as software testers or documentation maintainers, will be identified or credited as authors, as well as criteria for ordering the list of authors in cases of multiple authors

\item How the resource handles large numbers of authors and group or consortium authorship

\item Including guidelines about how changes to authorship are handled

\item What role the registry will play, if any, in authorship disputes, and if so, how they are handled

\item Maintaining consistency with the citation policies for the registry/repository

\item Using a credit ontology (\emph{e.g.}, \url{https://casrai.org/credit/}) to describe authors' contributions

\end{itemize}

\section{Best Practice: Share your metadata schema}
\label{best-practice-share-your-metadata-schema}

\textbf{Why we recommend this}: For individual and organizational users interested in the information in registries and repositories, revealing the metadata schema used for the entries helps users understand the structure and properties of the deposited information. The metadata structure helps to inform users how they might interact with or ingest records in the collection. A metadata schema mapped to other schemas and an API specification can improve the interoperability between registries and repositories.

\textbf{This should describe}:

\begin{itemize}
\item What schema is used (e.g., \href{https://codemeta.github.io/}{\emph{CodeMeta}}, \href{https://schema.org/}{\emph{Schema.org}}) and its version number if a published standard schema is used, or, if a custom schema is used, a description of the schema and/or a data dictionary

\item Where the metadata documentation or its official site can be found

\item What metadata is expected when submitting software, including which fields are required and which are optional, and the format of the content in each field

\end{itemize}

\textbf{What you might consider when stating your metadata schema:}

\begin{itemize}
\item Using a metadata schema that is mapped (``cross-walked'') to published standard schemas, or providing a cross-walk between your schema and other schemas

\item Providing an example of the metadata schema with a complete entry in your repository that illustrates all the fields of the schema

\end{itemize}

\section{Best Practice: Stipulate conditions of use}
\label{best-practice-stipulate-conditions-of-use}

\textbf{Why we recommend this}: A conditions of use policy lets users of your resource know how the metadata of the registry or repository can be used, attributed, and/or cited. It provides information about licensing and forestalls potential liabilities and difficulties that may arise, such as claims of damage for misinterpretation or misapplication of metadata. In turn, it clearly states how the metadata can and cannot be used, including for commercial purposes and in aggregate form.

\textbf{This should describe}:

\begin{itemize}
\item Legal disclaimers about the responsibility and liability borne by the registry or repository

\item License and copyright information, both for individual entries and for the registry or repository as a whole

\item Conditions for the use of the metadata, including prohibitions, if any

\item Preferred format for citing software entries

\item Preferred format for attributing or citing the resource itself

\end{itemize}

\textbf{What you might consider when writing a conditions of use policy:}

\begin{itemize}
\item What license governs your metadata, and whether there are licensing requirements for findings and/or derivatives of the resource

\item Whether there are differences in the terms and license for commercial versus noncommercial use

\item Conditions for the use of the API if one is available 

\item Restrictions on use of the metadata

\item Including a statement to the effect that the registry or repository makes no guarantees about completeness and is not liable for any damages that may arise from the use of the information

\end{itemize}

\section{Best Practice: State a privacy policy}
\label{best-practice-state-a-privacy-policy}

\textbf{Why we recommend this:} Having a privacy policy demonstrates a strong commitment to the privacy of users of the registry or repository, and allows the resource to comply with the legal requirement of many countries in addition to those a home institution and/or funding agencies may impose. A privacy policy discloses what information, analytics, and metrics a registry collects and/or retains about its users and why.

\textbf{This should describe}:

\begin{itemize}
\item What information is collected and how long it is retained

\item How the information, especially any personal data, is used

\item Whether tracking is done, what is tracked, and how (e.g., Google Analytics)

\item Whether cookies are used

\end{itemize}

\textbf{What you might consider when writing a privacy policy:}

\begin{itemize}
\item Detailing the specific data collected, why it is collected, and whether it is shared or sold

\item Being explicit about third party tools used to collect analytic information and potentially referencing their privacy policies

\item Stating whether users will receive email as a result of visiting or downloading content

\item Explaining the measures taken to protect users' privacy, and whether the resource complies with the \href{https://gdpr-info.eu/}{\emph{European Union Directive on General Data Protection Regulation}} (GDPR) or other local laws, if applicable

\item Reserving the right to make changes to the Privacy Policy

\item Defining a mechanism by which users can request information be removed

\end{itemize}

\section{Best Practice: Provide a retention policy}
\label{best-practice-provide-a-retention-policy}

\textbf{Why we recommend this}: Software registries and repositories make an implicit promise to retain records for some period of time, but for various reasons may have to remove records. Common examples include removing entries that are outdated or no longer meet the scope of the registry or are found to be in violation of policies. The collection should document retention goals so that users and depositors are aware of them.

\textbf{This should describe}:

\begin{itemize}
\item The length of time metadata and/or files are expected to be retained

\item Under what conditions metadata and/or files are removed

\item Who has the responsibility and ability to remove information

\item Procedures to request that metadata and/or files be removed

\end{itemize}

\textbf{What you might consider when writing a retention policy:}

\begin{itemize}
\item If assigning identifiers, whether best practices for persistent identifiers are followed, including resolvability, retention, and non-reuse of those identifiers

\item Making sure the length of time is not too prescriptive (e.g., ``for the next 10 years''), but rather fits within the context of the underlying organization(s) and its funding

\item Stating who is allowed to edit metadata, delete records, or delete files, and if so, how these changes are documented and consistent with the registry broadly

\item Explaining the process by which data may be taken offline and archived as well as the process for its possible retrieval

\end{itemize}

\section{Best Practice: Disclose your end-of-life policy}
\label{best-practice-disclose-your-end-of-life-policy}

\textbf{Why we recommend this}: Sharing a clear end-of-life policy increases trust in the community served by your registry or repository.  It demonstrates a thoughtful commitment to users by informing them that provisions for the artifacts contained in the resource have been considered should the resource close or otherwise end its services for these artifacts. Such a policy sets expectations and provides reassurance as to how long the records within the resource will be findable and accessible in the future.

\textbf{This should describe}:

\begin{itemize}
\item Under what circumstances the resource might end its services

\item What consequences would result from closure

\item What will happen to the metadata and/or the software artifacts contained in the resource in the event of closure

\item If long-term preservation is expected, where metadata and/or software artifacts will be migrated for preservation

\item How a migration will be funded

\end{itemize}

\textbf{What you might consider when writing a end-of-life policy:}

\begin{itemize}
\item Whether the records will remain available, and if so, how and for whom, and under which conditions, such as archived status or ``read only'', should the collection close

\item What restrictions, if any, may apply

\item Establishing a formal agreement or MOU with another registry, repository, or institution to receive and preserve the data or project, if applicable

\end{itemize}

\section{Policy examples}
\label{policy-examples}

\vspace*{-3pt}
\subsection{Scope Statement}
\label{scope-statement}

\begin{itemize}

\item Astrophysics Source Code Library. (n.d.). \emph{Editorial policy}.\\
\url{https://ascl.net/wordpress/submissions/editiorial-policy/}

\item bio.tools. (n.d.). \emph{Curators Guide}.\\
\url{https://biotools.readthedocs.io/en/latest/curators\_guide.html}

\item Caltech Library. (2017). \emph{Terms of Deposit}.\\
\url{https://data.caltech.edu/terms}

\item Caltech Library. (2019). \emph{CaltechDATA FAQ}.\\
\url{https://www.library.caltech.edu/caltechdata/faq}

\item Computational Infrastructure for Geodynamics. (n.d.). \emph{Code Donation}.\\
\url{https://geodynamics.org/cig/dev/code-donation/}

\item CoMSES Net Computational Model Library. (n.d.). \emph{Frequently Asked Questions}.\\
\url{https://www.comses.net/about/faq/\#model-library}

\item ORNL DAAC for Biogeochemical Dynamics. (n.d.). 
\emph{Data Scope and Acceptance Policy}.\\
\url{https://daac.ornl.gov/submit/}

\item RDA Registry and Research Data Australia. (2018). \emph{Collection}.
ARDC Intranet.\\
\url{https://intranet.ands.org.au/display/DOC/Collection}

\item Remote Sensing Code Library. (n.d.). \emph{Submit}.\\
\url{https://rscl-grss.org/submit.php}

\item SciCrunch. (n.d.). \emph{Curation Guide for SciCrunch Registry}.\\
\url{https://scicrunch.org/page/Curation\%20Guidelines}

\item U.S. Department of Energy: Office of Scientific and Technical Information. (n.d.-a). \emph{DOE CODE: Software Policy}.
\url{https://www.osti.gov/doecode/policy}

\item U.S. Department of Energy: Office of Scientific and Technical
Information. (n.d.-b). \emph{FAQs}. OSTI.GOV.\\
\url{https://www.osti.gov/faqs}

\end{itemize}

\vspace*{-2pt}
\subsection{Authorship}
\label{authorship}

\begin{itemize}

\item CASRAI. (n.d.). CRediT - Contributor Roles Taxonomy.\\
\url{https://casrai.org/credit/}

\item Committee on Publication Ethics: COPE. (2020a). \emph{Authorship and contributorship}.\\
\url{https://publicationethics.org/authorship}

\item Committee on Publication Ethics: COPE. (2020b). \emph{Core practices}.\\
\url{https://publicationethics.org/core-practices}

\item Dagstuhl EAS Specification Draft. (2016). \emph{The Software Credit Ontology}.\\
\url{https://dagstuhleas.github.io/SoftwareCreditRoles/doc/index-en.html\#}

\item Journal of Open Source Software. (n.d.). \emph{Ethics Guidelines}.\\
\url{https://joss.theoj.org/about\#ethics}

\item ORNL DAAC (n.d) \emph{Authorship Policy}.\\
\url{https://daac.ornl.gov/submit/}

\item PeerJ Journals. (n.d.-a). \emph{Author Policies}.\\
\url{https://peerj.com/about/policies-and-procedures/\#author-policies}

\item PeerJ Journals. (n.d.-b). \emph{Publication Ethics}.\\
\url{https://peerj.com/about/policies-and-procedures/\#publication-ethics}

\item PLOS ONE. (n.d.). \emph{Authorship}.\\
\url{https://journals.plos.org/plosone/s/authorship}

\item National Center for Data to Health. (2019). The Contributor Role Ontology.\\ 
\url{https://github.com/data2health/contributor-role-ontology}

\end{itemize}

\vspace*{-2pt}
\subsection{Metadata Schema}
\label{metadata-schema}

\begin{itemize}

\item ANDS: Australian National Data Service. (n.d.). \emph{Metadata}. ANDS.\\
\url{https://www.ands.org.au/working-with-data/metadata}

\item ANDS: Australian National Data Service. (2016). \emph{ANDS Guide: Metadata}.\\
\url{https://www.ands.org.au/__data/assets/pdf_file/0004/728041/Metadata-Workinglevel.pdf}

\item Bernal, I. (2019). \emph{Metadata for Data Repositories}.\\
\url{https://doi.org/10.5281/zenodo.3233486}

\item bio.tools. (2020). \emph{Bio-tools/biotoolsSchema} {[}HTML{]}.\\
\url{https://github.com/bio-tools/biotoolsSchema}
(Original work published 2015)

\item bio.tools. (2019). \emph{BiotoolsSchema documentation}.\\
\url{https://biotoolsschema.readthedocs.io/en/latest/}

\item The CodeMeta crosswalks. (n.d.)\\
\url{https://codemeta.github.io/crosswalk/}

\item Citation File Format (CFF). (n.d.)\\
\url{https://doi.org/10.5281/zenodo.1003149}

\item The DataVerse Project. (2020). DataVerse 4.0+ Metadata Crosswalk.\\
\url{https://docs.google.com/spreadsheets/d/10Luzti7svVTVKTA-px27oq3RxCUM-QbiTkm8iMd5C54}

\item OntoSoft. (2015). \emph{OntoSoft Ontology}.\\
\url{https://ontosoft.org/ontology/software/}

\item OpenAPI Specification. (2020).\\
\url{http://spec.openapis.org/oas/v3.0.3}

\item Zenodo. (n.d.-a). \emph{Schema for Depositing}.\\
\url{https://zenodo.org/schemas/records/record-v1.0.0.json}

\item Zenodo. (n.d.-b). \emph{Schema for Published Record}.\\
\url{https://zenodo.org/schemas/deposits/records/legacyrecord.json}

\end{itemize}

\vspace*{-2pt}
\subsection{Conditions of use policy}
\label{conditions-of-use-policy}

\begin{itemize}

\item Allen Institute. (n.d.). \emph{Terms of Use}.\\
\url{https://alleninstitute.org/legal/terms-use/}

\item Europeana. (n.d.). \emph{Usage Guidelines for Metadata}. Europeana Collections.\\
\url{https://www.europeana.eu/portal/en/rights/metadata.html}

\item U.S. Department of Energy: Office of Scientific and Technical Information. (n.d.). \emph{DOE CODE FAQ: Are there restrictions on the use of the material in DOE CODE?}\\
\url{https://www.osti.gov/doecode/faq\#are-there-restrictions}

\item Zenodo. (n.d.). \emph{Terms of Use}.\\
\url{https://about.zenodo.org/terms/}

\end{itemize}

\subsection{Privacy policy}
\label{privacy-policy}

\begin{itemize}

\item Allen Institute. (n.d.). \emph{Privacy Policy}.\\
\url{https://alleninstitute.org/legal/privacy-policy/}

\item CoMSES Net. (n.d.). \emph{Data Privacy Policy.\\
\url{https://www.comses.net/about/data-privacy/}}

\item Nature. (2020). \emph{Privacy Policy}.\\
\url{https://www.nature.com/info/privacy}

\item Research Data Australia. (n.d.). \emph{Privacy Policy}.\\
\url{https://researchdata.ands.org.au/page/privacy}

\item SciCrunch. (2018). \emph{Privacy Policy}. SciCrunch.\\
\url{https://scicrunch.org/page/privacy}

\item Science Repository. (n.d.). \emph{Privacy Policies}.\\
\url{https://www.sciencerepository.org/privacy}

\item Zenodo. (n.d.). \emph{Privacy policy}.\\
\url{https://about.zenodo.org/privacy-policy/}

\end{itemize}

\subsection{Retention Policy}
\label{retention-policy}

\begin{itemize}

\item Caltech Library. (n.d.). \emph{CaltechDATA FAQ}.\\
\url{https://www.library.caltech.edu/caltechdata/faq}

\item CoMSES Net Computational Model Library. (n.d.). \emph{How long will models be stored in the Computational Model Library?\\
\url{https://www.comses.net/about/faq/}}

\item Dryad. (2020). \emph{Dryad FAQ - Publish and Preserve your Data}.\\
\url{https://datadryad.org/stash/faq\#preserved}

\item Software Heritage. (n.d.). \emph{Content policy}.\\
\url{https://www.softwareheritage.org/legal/content-policy/}

\item Zenodo. (n.d.). \emph{General Policies v1.0}.\\
\url{https://about.zenodo.org/policies/}

\item Bioconductor. (2020). \emph{Package End of Life Policy.\\
\url{https://bioconductor.org/developers/package-end-of-life/} }

\end{itemize}

\subsection{End-of-life policy}
\label{end-of-life-policy}

\begin{itemize}

\item Figshare. (n.d.). \emph{Preservation and Continuity of Access Policy}.\\
\url{https://knowledge.figshare.com/articles/item/preservation-and-continuity-of-access-policy}

\item Open Science Framework. (2019). \emph{FAQs}. OSF Guides.\\
\url{http://help.osf.io/hc/en-us/articles/360019737894-FAQs}

\item NASA Earth Science Data Preservation Content Specification (n.d.)\\
\url{https://earthdata.nasa.gov/esdis/eso/standards-and-references/preservation-content-spec}

\item Zenodo. (n.d.). \emph{Frequently Asked Questions}.\\
\url{https://help.zenodo.org/}

\end{itemize}

\section{Additional useful sites }
\label{additional-useful-sites}

In addition to the links to sites and information embedded in this Concise Guide, the following sites are directly applicable to the best practices we have listed.

\begin{itemize}

\item (Authorship) Citation File Format: \url{https://citation-file-format.github.io/}

\item (Authorship) CiteAs: \url{https://citeas.org/}

\item (Metadata Schema) Software Heritage Metadata workflow: \url{https://docs.softwareheritage.org/devel/swh-indexer/metadata-workflow.html}

\item (Metadata Schema) W3C data profile definition: \url{https://www.w3.org/TR/dx-prof-conneg/\#dfn-data-profile}

\end{itemize}

You may also be interested in the \url{https://www.coretrustseal.org/why-certification/requirements/}, which are intended to reflect the characteristics of trustworthy repositories.

\section{Glossary}
\label{glossary}

\textbf{API}: \href{https://en.wikipedia.org/wiki/API}{Application Programming Interface}

\textbf{Collection}: Used in this document as a synonym for \emph{registries and repositories}

\textbf{Depositor}: A user who submits information and/or software to a registry or repository; synonymous with \emph{software contributor}

\textbf{Entry}: Information about and/or software for a particular holding in a registry or repository; synonymous with \emph{record}

\textbf{JSON-LD}: \href{https://en.wikipedia.org/wiki/JSON-LD}{JavaScript Object Notation for Linked Data}

\textbf{Metadata}: \href{https://en.wikipedia.org/wiki/Metadata}{Information about a code or software package}

\textbf{Record}: Information about and/or software for a particular holding in a registry or repository; synonymous with \emph{entry}

\textbf{Registry}: Typically an index or catalog of software stored elsewhere

\textbf{Repository}: Typically a site that both indexes and stores software

\textbf{Resource}: Used in this document as a synonym for \emph{registries and repositories}

\textbf{Software author}: A person who is credited as an author of a software package; this may include not only one who writes code, but also one who tests, documents, maintains, or otherwise contributes effort to the software package

\textbf{Software contributor}: A user who submits information and/or software to a registry or repository; synonymous with \emph{depositor}

\textbf{XML}: \href{https://en.wikipedia.org/wiki/XML}{Extensible Markup Language}

\section{Authors}
\label{authors}

% Tighten paragraph spacing to get everyone on one page.
\begin{spacing}{0.86}

\href{https://orcid.org/0000-0003-3150-4837}{Alain Monteil}, INRIA, \href{https://hal.archives-ouvertes.fr/}{HAL}/\href{https://www.softwareheritage.org/}{Software Heritage}

\href{https://orcid.org/0000-0003-3499-8262}{Alejandra Gonzalez-Beltran}, Science and Technology Facilities Council, UK Research and Innovation

\href{https://orcid.org/0000-0002-5082-6404}{Alexandros Ioannidis}, CERN, \href{https://zenodo.org/}{Zenodo}

\href{https://orcid.org/0000-0003-3477-2845}{Alice Allen}, University of Maryland, \href{http://ascl.net/}{Astrophysics Source Code Library}

\href{https://orcid.org/0000-0002-6523-6079}{Allen Lee}, Arizona State University, \href{https://www.comses.net/}{CoMSES Net}

\href{https://orcid.org/0000-0002-5497-0243}{Anita Bandrowski}, UCSD, \href{https://scicrunch.org/}{SciCrunch}

\href{https://orcid.org/0000-0002-1421-1728}{Bruce E. Wilson}, Oak Ridge National Laboratory, \href{https://daac.ornl.gov/}{ORNL Distributed Active Archive Center for Biogeochemical Dynamics}

\href{https://orcid.org/0000-0002-0381-3766}{Bryce Mecum}, NCEAS, UC Santa Barbara, \href{https://codemeta.github.io/}{CodeMeta}

\href{https://orcid.org/0000-0003-2538-607X}{Cai Fan Du}, iSchool, University of Texas at Austin, \href{http://citeas.org/}{CiteAs}

\href{https://orcid.org/0000-0002-8523-1478}{Carly Robinson}, \href{https://www.osti.gov/}{DOE-OSTI}

\href{http://orcid.org/0000-0003-0454-7145}{Daniel Garijo}, Information Sciences Institute, University of Southern
California, \href{http://www.ontosoft.org/}{Ontosoft}

\href{https://orcid.org/0000-0001-5934-7525}{Daniel S. Katz}, University of Illinois at Urbana-Champaign, Associate
EiC for JOSS, \href{https://www.force11.org/group/software-citation-implementation-working-group}{FORCE11 Software Citation Implementation Working Group} co-chair

\href{https://orcid.org/0000-0002-1852-3972}{David Long}, Brigham Young University, \href{https://rscl-grss.org/}{IEEE GRS Remote Sensing Code Library}

\href{https://orcid.org/0000-0002-3057-0659}{Genevieve Milliken}, NYU Bobst Library, \href{https://investigating-archiving-git.gitlab.io/}{IASGE}

\href{https://orcid.org/0000-0002-7552-1009}{Hervé Ménager}, Institut Pasteur, \href{https://bio.tools/}{ELIXIR bio.tools}

\href{https://orcid.org/0000-0002-1861-1526}{Jessica Hausman}, Jet Propulsion Laboratory, \href{https://podaac.jpl.nasa.gov/}{PO.DAAC}

\href{https://orcid.org/0000-0002-7064-4069}{Jurriaan H. Spaaks}, Netherlands eScience Center, \href{https://www.research-software.nl/}{Research Software Directory}

\href{https://orcid.org/0000-0003-1483-5335}{Katrina Fenlon}, University of Maryland, \href{https://ischool.umd.edu/}{iSchool}

\href{https://orcid.org/0000-0003-1439-2204}{Kristin Vanderbilt}, Environmental Data Initiative, \href{https://imcr-hackathon.github.io/website/}{IMCR}

\href{http://orcid.org/0000-0002-1021-3101}{Lorraine Hwang}, \href{https://geodynamics.org/}{Computational Infrastructure for Geodynamics}, UC Davis

\href{https://orcid.org/0000-0002-4670-0964}{Lynn Davis}, DOE-OSTI

\href{https://orcid.org/0000-0003-1419-2405}{Martin Fenner}, DataCite, \href{https://www.force11.org/group/software-citation-implementation-working-group}{FORCE11 Software Citation Implementation Working Group} co-chair

\href{https://orcid.org/0000-0002-2961-9670}{Michael R. Crusoe}, CWL, \href{https://www.debian.org/devel/debian-med/}{Debian-Med}

\href{https://orcid.org/0000-0001-9105-5960}{Mike Hucka}, Caltech, \href{http://sbml.org/Main_Page}{SBML}, \href{http://co.mbine.org}{COMBINE}

\href{https://orcid.org/0000-0003-1206-3431}{Mingfang Wu}, \href{https://ardc.edu.au/}{Australian Research Data Commons}

\href{https://orcid.org/0000-0002-8876-7606}{Neil Chue Hong}, \href{https://www.software.ac.uk/}{Software Sustainability Institute}, University of Edinburgh, \href{https://www.force11.org/group/software-citation-implementation-working-group}{FORCE11 Software Citation Implementation Working Group} co-chair

\href{https://orcid.org/0000-0003-1774-3436}{Peter Teuben}, University of Maryland

\href{https://orcid.org/0000-0003-2926-8353}{Shelley Stall}, American Geophysical Union, \href{https://www.agu.org/Learn-About-AGU/About-AGU/Data-Leadership}{AGU Data Services}

\href{https://orcid.org/0000-0003-4925-7248}{Stephan Druskat}, German Aerospace Center (DLR)/University
Jena/Humboldt-Universität zu Berlin, \href{https://citation-file-format.github.io/}{Citation File Format}

Ted Carnevale, Neuroscience Department, Yale University, \href{https://senselab.med.yale.edu/modeldb/}{ModelDB}

\href{https://orcid.org/0000-0001-9266-5146}{Tom Morrell}, Caltech, \href{https://data.caltech.edu/}{CaltechDATA}

\end{spacing}

\end{document}